\documentclass[12pt]{article}
\usepackage[round,colon]{natbib}
\usepackage{amsmath}
\usepackage{epsfig}

\normalsize

\begin{document}

\title{Biological activity in the wake of an island close to a
coastal upwelling}

\vfill

\author{ Mathias Sandulescu$^1$,
Crist\'obal L{\'o}pez$^2$,\\
 Emilio
Hern{\'a}ndez-Garc{\'\i}a$^2$,
and Ulrike Feudel$^1$ \\
\  \\
 $^1$ Carl-von-Ossietzky Universit\"at Oldenburg \\
D-26111
Oldenburg, Germany \\
$^2$ IFISC, Instituto de F\'\i sica Interdisciplinar y Sistemas
Complejos \\
 (CSIC - Universitat de les Illes Balears) \\
E-07122 Palma de Mallorca, Spain \ \\ \ } \vfill \maketitle


\begin{abstract}
Hydrodynamic forcing plays an important role in shaping the
dynamics of marine organisms, in particular of plankton. In this
work we study the planktonic biological activity in the wake of an
island which is close to an upwelling region. Our research is
based on numerical analysis of a kinematic flow mimicking the
hydrodynamics in the wake, coupled to a three component plankton
model.

We use parameter values of relevance for the Canary wake, and the
main results for a realistic range of parameters in this area area
are: a) Primary production is enhanced in the region of the wake
opposite to the upwelling zone. b) There is a strong dependence of
the productivity on the inflow conditions of biological material
entering the wake transported by the main current. Finally c) we
show that under certain conditions the interplay between wake
structures and biological growth leads to plankton blooms inside
mesoscale hydrodynamic vortices that act as incubators of primary
production.


\end{abstract}



\newpage

\section{Introduction}

Understanding the influence of hydrodynamic motions on the growth,
productivity and distribution of marine organisms, especially in
the context of plankton dynamics, is a major challenge recently
addressed from a variety of perspectives
\citep{Mann1991,Denman1995,Abraham1998,Peters2000,Karolyi2000,Lopez2001b,Lopez2001,Martin2002,Martin2003}.
Vertical transport processes of nutrients are recognized as key
factors controlling plankton productivity \citep{Denman1995}. In
particular, upwelling areas in the world's oceans are of
fundamental importance for the growth of phytoplankton which is
the base of oceanic food webs. They are characterized by nutrient
rich waters coming to the surface from depths of over $50$ meters.
Nutrient enrichment enhances phytoplankton growth close to the
upwelling regions, giving rise to an increase in zooplankton and
fish populations in the area. More recently, the importance of
horizontal fluid motion has also been pointed out
\citep{Abraham1998,Lopez2001,Hernandez2002,Hernandez2003b,Martin2003}.
Mesoscale stirring redistributes and mixes plankton and nutrients
laterally, giving rise also to enhanced productivity
\citep{Martin2002}, or to bloom initiation \citep{Reigada2003},
and affects species competition and coexistence
\citep{Karolyi2000,Bracco2000}. Satellite images illustrate the
interaction between horizontal mesoscale motions and plankton
dynamics.

Vertical upwelling and strong mesoscale activity occur
simultaneously in several places of the globe. A stronger impact
and a high complexity of the physical-biological interactions are
expected there. Some of these areas are the Benguela zone,  the
Humboldt Current, or the Canary islands.


Though the phenomena we discuss are rather general, we illustrate
them by using the Canary islands, which are close to the
northwestern African coast, as a specific example. There,
upwelling occurs at the African coast because of Ekman pumping
induced by the dominant winds, and in addition, the Canary islands
constitute an obstacle for the main ocean current in the area,
flowing from Northeast to Southwest, originating a strong
mesoscale hydrodynamic activity in their wake. The interaction
between the vortices in the wake and the Ekman flow transporting
nutrient-rich waters from the coastal upwelling seems to be at the
heart of the observed enhancement of biological production in the
open Atlantic ocean close to the Canary region. Motivated by this
situation, the aim of this paper is to study, in a more general
framework, the role of wake vorticity in redistributing upwelled
nutrients and influencing phytoplankton growth.

To this end we combine the kinematic model flow introduced in
\citet{Sandulescu2006} with a simple model of a
Nutrient-Phytoplankton-Zooplankton (NPZ) trophic chain, and study
the impact of the flow characteristics on the biological dynamics,
particularly on the primary production ($PP$). We will use mainly
parameter values of relevance in modeling the Canary wake, but we
expect our results to have broader application. Only horizontal
transport is explicitly taken into account in the flow, the
upwelling is modelled as a source term in the nutrient equation.
We address questions such as (i) whether the island wake is a
barrier for the upwelled nutrients, or (ii) if rather the
generated stirring mixes nutrients into poorer waters so that
primary production is enhanced, or (iii) what is  the impact of
the presence of vortices and other wake structures on biological
activity. Our main results are, on the one hand, that for a range
of parameters  which is realistic in the Canary area, primary
production is enhanced in the part of the wake opposite to the
upwelling zone. That is, the wake is not a barrier confining the
region of high nutrients and plankton growth. On the other hand,
there is a strong dependence of the productivity, and of the role
of the vortices, on the inflow of biological components entering
the wake due to transport by the main current. In some situations
the vortices in the wake act as an {\sl incubator} whose sole
presence is enough to greatly enhance biological productivity in
poor waters entering the region.

Recently, in \citet{Sandulescu2007} we have studied the same model
but attending to the role of the different time scales present in
the system. We find there that the long residence times of
nutrients and plankton close to the island, and the confinement of
plankton inside vortices are important factors for the appearence
of localized plankton blooms. These studies complement the final
part of this paper where we focus on the role of vortices on
biological production.

In the next section we present our general modeling framework,
presenting the velocity field, the plankton model, and the
boundary (environmental) conditions. In section \ref{results} we
present our results, organized in two subsections that contain our
studies of concentrations of nutrients, plankton, and primary
production in relation to hydrodynamic and inflow conditions, and
another one in which we analyze the plankton content of vortices.
Section \ref{conclusions} summarizes our conclusions.

\section{Modeling framework}
\label{general}

\begin{figure}
\centering{ \mbox{
\includegraphics[angle=270, width=\textwidth]{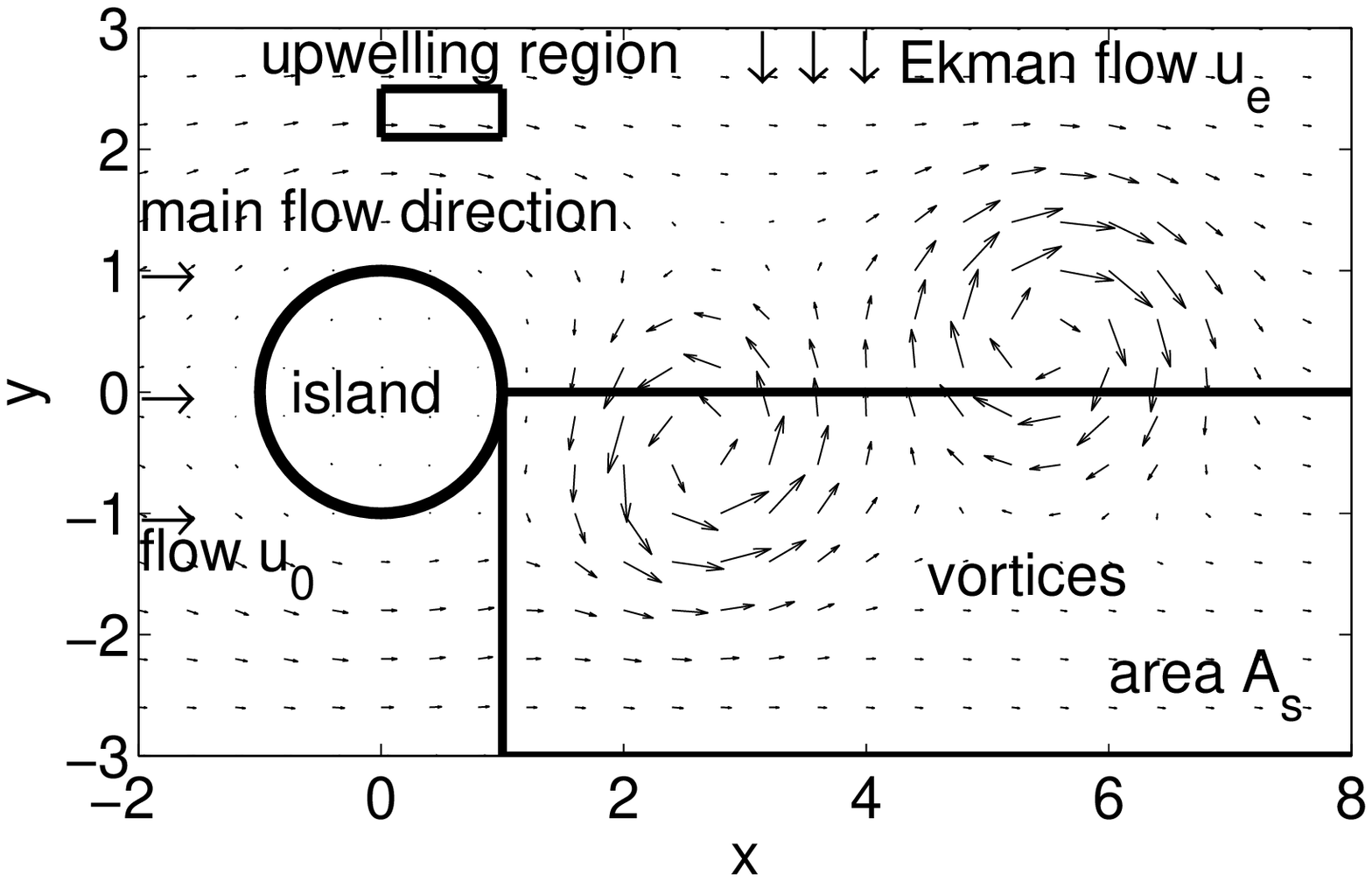}
} }\caption{The computational domain with a snapshot of the
velocity field. Spatial coordinates, $x$  and $y$, are in units of
the island radius.
\label{fig:area}}
\end{figure}

Figure \ref{fig:area} shows our two-dimensional model domain. The
main current flows from left to right, passing by the circular
obstacle, which models the presence of an island, and giving rise
to mesoscale vortices in its wake. Vertical hydrodynamic motion is
not explicitly considered, but its effect on nutrient upwelling is
modelled by a source of nutrients (shown as a small box in the
upper zone of the domain). The associated Ekman flow points
towards the interior of the domain. Our focus of study will be the
lower part of the wake, the region $A_s$, marked with a box in the
lower part of figure \ref{fig:area}. A discussion about the choice
of a different $A_s$ and its negligible effect in our results can
be consulted in \cite{Sandulescu2006}. Our primary objective is to
determine if nutrient input from the upwelling region, which is on
the opposite part of the wake, may enhance the biological activity
in this region, and to elucidate the role, as barriers or as
transporters, of the wake and of the vortices present in it. In
addressing this goal, we realize the importance of the contents of
the water transported towards our domain by the main flow.
 In the context of
the Canary islands situation, that will guide our selection of
parameter values, the upper part of figure \ref{fig:area}
represents the African coast, with the coastal upwelling. The
obstacle is the Canary archipelago, more particularly the Gran
Canaria island which plays an important role for the emergence of
the vortices in the area. The main current is the Canary current,
flowing from northeast to southwest. In this context our analysis
may be of relevance to discuss enrichment of the open ocean beyond
the Canary wake by input of coastal waters. More generally, it
illustrates the interplay between transport, stirring, and
biological dynamics.


\subsection{The velocity field}
\label{The velocity field}

We briefly introduce the velocity field used in this study. It is
essentially the horizontal incompressible flow, derived from a
time periodic streamfunction of period $T_c$, proposed by
\citet{Jung1993} to model kinematically the vortex street behind a
cylinder at moderate Reynolds numbers, but modified to include a
velocity component pointing towards the domain interior that
mimics the Ekman flow associated with the upwelling
\citep{Sandulescu2006}. Its technical description
 can be found in \citet{Sandulescu2006}.
 We next describe
it qualitatively.

There are a maximum of two vortices simultaneously in the system.
They are of opposite vorticity sign but their maximal vortex
strength denoted by $w$ is equal. They are created behind the
circular obstacle with a phase difference of half a period,
$T_c/2$. Each of the vortices travels a distance along the $x$
direction for a time $T_{c}$ and finally disappears. Then the
process repeats periodically again. Since real oceanic flows are
never perfectly periodic we add some randomness to the vortex
trajectories. Instead of moving along straight horizontal lines,
they are subjected to a stochastic transversal Brownian motion of
small amplitude (see details in \citet{Sandulescu2006}).

The main background flow moves in the positive horizontal
direction with a speed $u_0$, and the Ekman drift, which is
intended to model the flow from the coast towards the ocean
interior, is introduced by considering an additional velocity
$u_E$ in the $y$ direction acting in the region with $x$
coordinate larger than the island radius $r$, i.e. just behind the
island ($x \geq r$) (this additional component $u_E$ is indicated
in \ref{fig:area}, but not added to the velocity field plotted
there). The circular obstacle, which is considered as a model
island, has a radius $r$.

To adapt this general setup of the velocity field to a realistic
and more specific situation we choose parameter values which are
guided by the values in the Canary wake \citep{Sandulescu2006}:
$r=25$ km, $u_0=0.18$ m/s, $u_E=0.02$ m/s, and $T_c=30$ days. We
consider two situations for the vortex strength in the wake.
Previous results \citep{Sandulescu2006} indicate that the wake
entrains water from one side of the island towards the other in
form of filaments for high values of the vortex strength which are
realistic in the Canary area ($w=w_H \approx 55\times10^3$
m$^2$/s), but that it acts as a barrier to transport when the
vortices are weak (say $w =w_L\approx w_H/20 = 2.75\times10^3$
m$^2$/s).
 We will analyze the
plankton dynamics under these two vortex strengths, $w_L$ and
$w_H$, and also at intermediate ones. Despite the smallness of
$u_E$, it is larger than the minimum needed to observe a
transition from no transport to transport when increasing
$w$~\citep{Sandulescu2006}.

\subsection{The NPZ model}

Our description of the plankton population dynamics is based on a
model developed by \cite{Oschlies1999}. It describes the
interaction of a three level trophic chain in the mixed layer of
the ocean, consisting of nutrients $N$, phytoplankton $P$ and
zooplankton $Z$, whose concentrations evolve in time with the
following $NPZ$ dynamics:

\begin{eqnarray}
\frac{dN}{dt} &=& F_N \equiv \Phi_{N} - \beta \frac{N}{k_{N}+N}P \nonumber\\
&+& \mu_{N} \left((1-\gamma) \frac{\alpha \eta P^{2}}{\alpha +
\eta P^{2}}Z + \mu_{P}P + \mu_{Z}Z^{2}\right), \nonumber
\\
\frac{dP}{dt} &=& F_P \equiv \beta \frac{N}{k_{N}+N}P -
\frac{\alpha \eta P^{2}}{\alpha+ \eta P^{2}}Z - \mu_{P}P,
\nonumber
\\
\frac{dZ}{dt} &=& F_Z \equiv \gamma \frac{\alpha \eta
P^{2}}{\alpha + \eta P^{2}}Z - \mu_{Z}Z^{2}.
\label{BioDLG}
\end{eqnarray}

 The dynamics
of the nutrients includes three different processes. There is a
nutrient supply given by $\Phi_{N}=S(x,y)(N_0-N)$ due to vertical
mixing. $S$ gives the inverse of the time scale for the nutrients
to relax to the nutrient concentration $N_0$ below the mixed
layer. Therefore $S$ is the parameter accounting for the vertical
nutrient supply due to upwelling. We take $S(x,y)=S_l=0.00648 \
day^{-1}$ outside of the upwelling region and
$S(x,y)=S_h=100S_l=0.648 \ day^{-1}$ in the nutrient-rich
upwelling area identified in figure \ref{fig:area}. The nutrients
are consumed by the phytoplankton according to a Holling type II
functional response. The last three terms inside the parenthesis
of the nutrient equation denote the recycling of a part of all
dead organic matter.
The phytoplankton grows upon the consumption of the nutrients, but
its concentration is decreased due to grazing by zooplankton and
to natural mortality. The grazing
 enters as a growth term
for the zooplankton concentration with an efficiency factor
$\gamma$. Zooplankton mortality is assumed to be quadratic.
Additional details can be consulted in \citet{Oschlies1999} and
\citet{Pasquero2005}.
 The parameters used are taken from
\cite{Pasquero2004} and presented in the Table
\ref{table:bioparameters}.

\begin{table}   
\begin{tabular}[t]{l l}
 \hline
parameter & value \\
 \hline
$\beta$ &      0.66 day$^{-1}$ \\
$\eta$ &   1.0 (mmol~N~m$^{-3})^{-2}$~day$^{-1}$ \\
$\gamma$ &      0.75 \\
$\alpha$ &      2.0~day$^{-1}$ \\
$S_l$ &         0.00648 day$^{-1}$ (nutrient poor) \\
$S_h$ &         0.648 day$^{-1}$ (nutrient rich) \\
$k_N$ &         0.5 mmol N m$^{-3}$\\
$\mu_N$ &       0.2 \\
$\mu_P$ &       0.03 day $^{-1}$ \\
$\mu_Z$ &       0.2 (mmol~N~m$^{-3})^{-1}$~day$^{-1}$ \\
$N_0$ &         8.0 mmol~N~m$^{-3}$ \\
 \hline
\end{tabular}
\caption{List of parameters used in the biological model}
\label{table:bioparameters}
\end{table}

The primary production, the rate at which new organic matter is
produced, is given by the growth term in the phytoplankton
dynamics
\begin{equation}\label{PP}
PP=\beta \frac{N}{k_{N}+N}P.
\end{equation}

The dynamics of this food chain model is studied in detail in
\cite{Edwards1996} and \cite{Pasquero2004}. Depending on the
parameters of the model, it exhibits stationary or oscillatory
behavior in the long-term limit. The chosen parameter values lead
to a steady state. Using the values from Table
\ref{table:bioparameters} and fixing the vertical mixing to the
lower value $S=S_l=0.00648$ day$^{-1}$ we obtain a steady state
that will be called the {\sl ambient} state: $N_{amb}=0.185$,
$P_{amb}=0.355$ and $Z_{amb}=0.444$ mmol~N~m$^{-3}$. Thus, in the
nutrient poor region occupying most of the domain the ambient
primary production in steady state is $PP_{amb}= 0.0633$
mmol~N~m$^{-3}$ day$^{-1}$. In the upwelling region, $S=S_h=0.648$
day$^{-1}$ and the steady state that would be reached under this
nutrient input would be $N_{up}=7.539$, $P_{up}=0.603$ and
$Z_{up}=1.154$ mmol~N~m$^{-3}$, and the primary production
associated with these values would be $PP_{up}=0.373$
mmol~N~m$^{-3}$, i.e. nearly 6 times $PP_{amb}$.

\subsection{Complete model and input conditions}
\label{algorithm}

The coupling of the biological and the hydrodynamic model yields
an advection-reaction system. We add also an eddy diffusion
process acting on plankton and nutrients concentrations with
diffusion coefficient $D$ to incorporate the small scale
turbulence, which is not explicitly taken into account by the
large scale velocity field used. Following \citet{Okubo1971}
prescriptions, we take $D \approx 10 m^2/s$, corresponding to
spatial scales of about $10~km$ at which flow details begin to be
absent from our large scale flow model.
 Thus our complete model is
given by the partial differential equations:
\begin{eqnarray}
\frac{\partial N}{\partial t} + {\bf v}\cdot \nabla N &=& F_N + D
\nabla^2 N, \nonumber
\\
\frac{\partial P}{\partial t} + {\bf v}\cdot \nabla P &=& F_P + D
\nabla^2 P , \nonumber
\\
\frac{\partial Z}{\partial t} + {\bf v}\cdot \nabla Z &=& F_Z + D
\nabla^2 Z ,
\label{PDE}
\end{eqnarray}
with the biological interactions $F_N$, $F_P$, and $F_Z$ from Eq.
(\ref{BioDLG}), and the velocity field ${\bf v} (x,y,t)$ described
in subsection \ref{The velocity field}. This system is numerically
solved by means of a semi-Lagrangian algorithm on a grid.
Additional details of the integration algorithm are  reported in
the Appendix~\ref{appendix:algorithm}.

Since we are studying an open flow, inflow conditions into the
left part of the domain should be specified.
 It turns out that the influence of
inflow concentrations is rather important and we present here two
cases that exemplify the two main behaviors we have identified:
 In the first one fluid parcels enter
the computational domain with the ambient concentrations
$N_{amb}$, $P_{amb}$, and $Z_{amb}$. This corresponds to the
steady state for $S=S_l$, and represents the situation in which
the exterior of the computational domain has the same properties
as the part of the domain without upwelling. This input condition
will allow us to focus on the interaction between the upwelling
water and the main part of the domain containing the wake. In our
second situation fluid particles transported by the main flow
enter the domain from the left with a close to vanishing  content
of nutrients and plankton, corresponding to a biologically very
poor open ocean outside the considered domain. To be specific, we
take $N_L=0.01 N_{amb}$, $P_L=0.01 P_{amb}$ and $Z_L=0.01
Z_{amb}$. Primary production in the inflow water is very low:
$PP_L=8.6\times 10^{-6}$ mmol~N~m$^{-3}$ day$^{-1}$. This is more
than 7000 times smaller than $PP_{amb}$. Since those
concentrations are very low, we take into account that
fluctuations may be important by adding to each of the
concentrations $(N_L,P_L,Z_L)$ of each fluid parcel entering the
system an independent random amount
of about $\sim$ 5\% of the inflow concentration. In this second
situation there is mixing between three types of water: the
`ambient', the `upwelled', and the 'inflow' ones. It turns out
that the interaction between inflow and wake will be the
responsible for the interesting behavior described below.

\begin{figure} 
\centering{\mbox{
\includegraphics [angle=0,width=\textwidth]{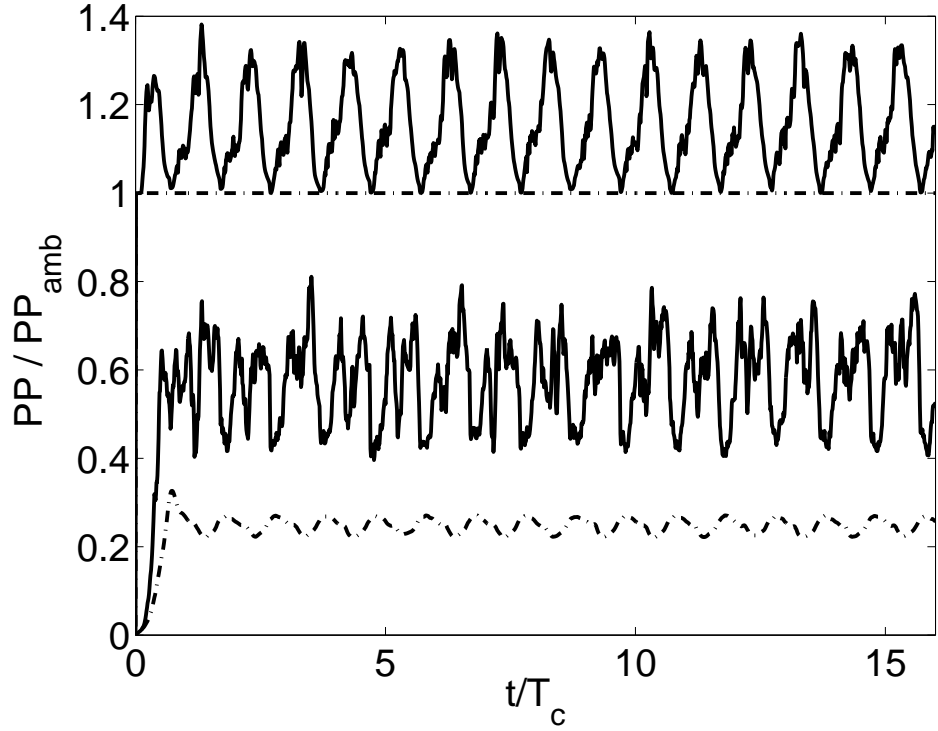}
} }
\caption{The time evolution of the ratio between
the primary production, $PP$, spatially averaged in $A_s$, and the
ambient one $PP_{amb}$ for the two inflow cases and two values of
the vortex strength. The two upper lines are for ambient input
concentrations, and the two lower for low inflow. Dashed-dotted
lines $w=w_L$, solid lines $w=w_H$.
\label{fig:PP-over-t}}
\end{figure}

\section{Results}
\label{results}

In this section we first describe the outcome of the different
scenarios considered in terms of primary production and plankton
distributions, and then address in more detail the relation
between vortex structures and plankton patches. The brackets $<>$
shall denote spatiotemporal averages.

\subsection{Primary production and plankton dynamics}
\label{subs:PrimaryProduction}

We stress that one of the main observations in
\citet{Sandulescu2006} is that, for the flow parameters used here,
there is a qualitative change in the transport behavior at vortex
strength $w_c \approx w_H/10$: For weaker vortices, a plume of
passive tracers released from the location of our upwelling area
develops in the direction of the main flow with a slight
transverse displacement due to the Ekman flow $u_E$ but remaining
far from our study region $A_s$. The wake acts here as a {\sl
barrier} to transport. For $w>w_c$, however, the plume becomes a
filament that is entrained by the vortices, so that it crosses the
wake and reaches $A_s$. Note that $w_L < w_c <w_H$.

\begin{figure}
\centering{
\includegraphics [angle=0, width=\textwidth]{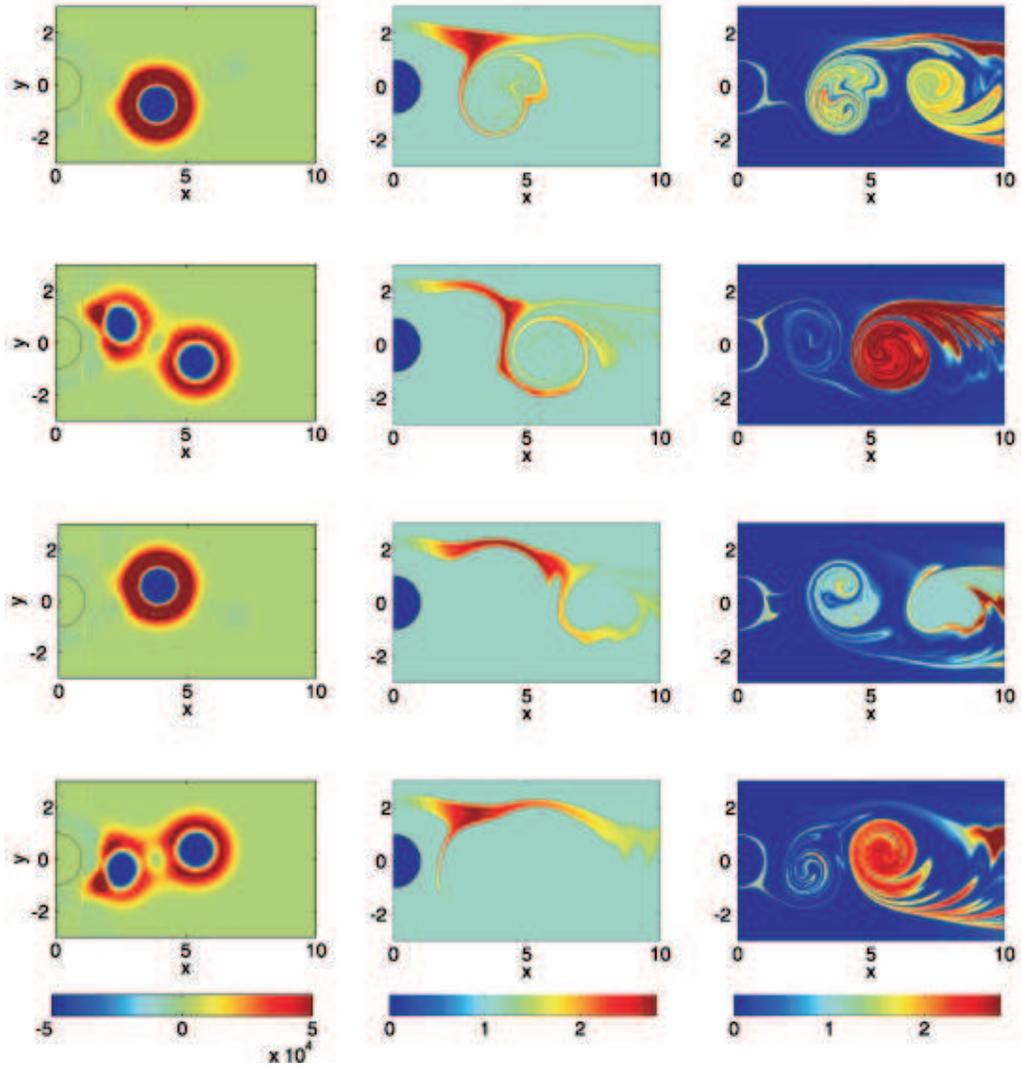}
}\caption{The Okubo-Weiss parameter (left column), the
concentration of phytoplankton for ambient inflow (middle), and
concentration of phytoplankton for the low inflow situation
(right). Phytoplankton concentration is expressed in units of
$P_{amb}$, and $w=w_H$. Snapshots taken during one flow period, at
$t/T_c= 14, 14.25, 14.5, 14.75$, from top to bottom.
\label{fig:patterns}}
\end{figure}

The observed behavior of our biological model when ambient
concentrations are used at the inflow reflects directly this
transport behavior: The two upper lines in figure
\ref{fig:PP-over-t} show the time evolution of the productivity
averaged over the region $A_s$. The dashed-dotted line with a
nearly constant value $PP/PP_{amb}\approx 1$ is obtained for
$w=w_L < w_c$. The upwelling plume fertilizes the upper part of
the computational domain, where higher concentrations of plankton
are observed, but the lower part of the wake is unaffected by this
and keeps its low ambient productivity value $PP_{amb}$ nearly
constant. When $w=w_H>w_c$ (upper solid line) productivity becomes
enhanced with respect to its ambient value. It undergoes roughly
periodic oscillations reflecting the periodic motion of the
nutrient filament entrained by the vortices. The central column in
figure \ref{fig:patterns} displays the phytoplankton spatial
distribution at different time instants. A filament of high
phytoplankton concentration appears in the system, sitting
basically on top of the high nutrient filament (not shown)
emerging from the upwelling and being entrained by the vortices.
Zooplankton and primary production are also distributed in the
same way.

\begin{figure}
\centering{ \mbox{
\includegraphics
[angle=0,width=\textwidth]{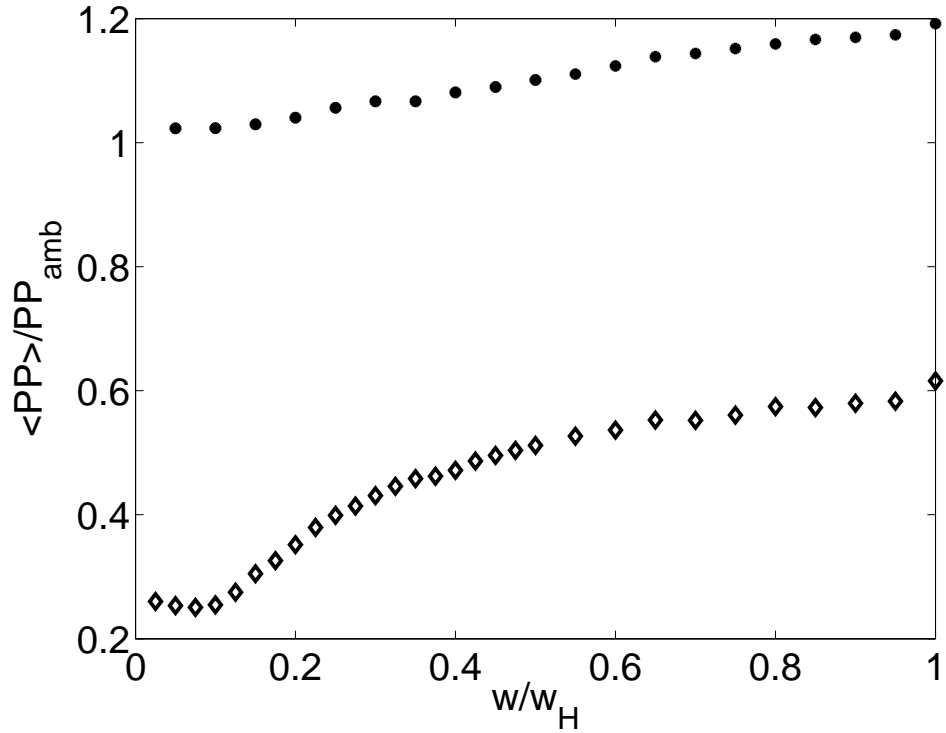} } }
\caption{The spatiotemporal average, $<PP>$, of the primary production in
$A_s$ in terms of its ambient value, $PP_{amb}$, as a function of
the vortex strength $w$ normalized by $w_H$. We plot the two
inflow cases: ambient inflow concentrations ($\bullet$) and low
inflow concentrations ($\diamond$). }
\label{fig:PP-over-w}
\end{figure}

Figure \ref{fig:PP-over-w} shows the average primary production
(averaged over $A_s$ and then averaged in time) as a function of
the vortex strength $w$ in the range $[0.025w_H, w_H]$. As
anticipated, a transition from essentially no enrichment by the
upwelling to an increasing primary production occurs around
$w=w_c\approx 0.1 w_H$, confirming a direct influence of the
physical transport process on the biological dynamics.

The dynamics in the low concentration inflow case is very
different. For all values of $w$ considered, the average primary
production in $A_s$ is smaller than the ambient one. This can be
understood from the fact that fluid elements enter the domain with
very low nutrient and plankton concentrations.

\begin{figure}
\centering{
\includegraphics [angle=0, width=\textwidth]{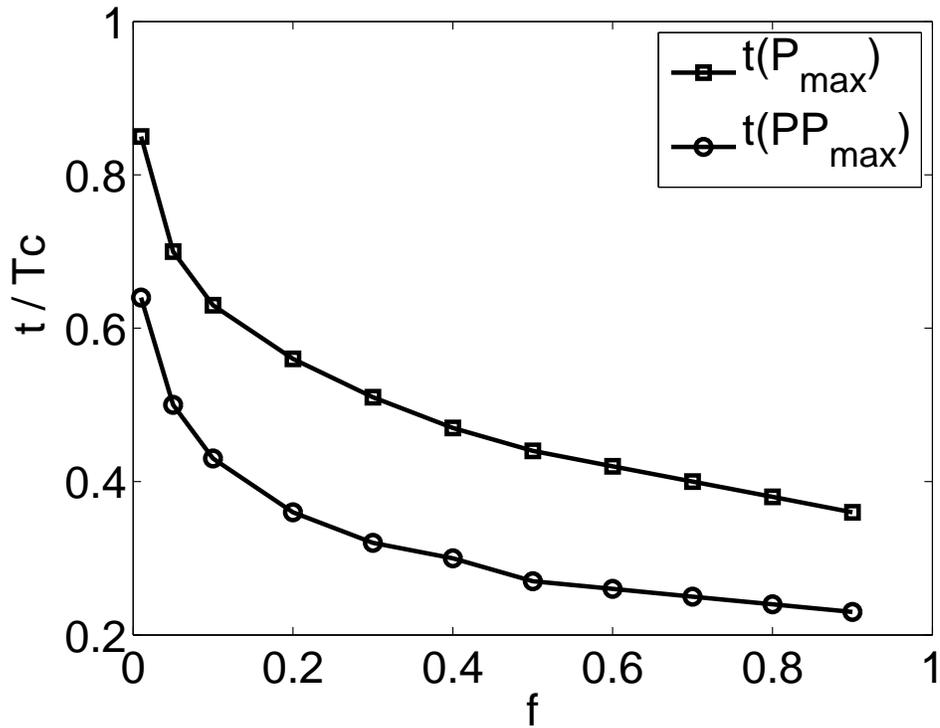}
}\caption{Time needed, in units of $T_c$,
 to reach the maximum value of $P$ or of $PP$, vs $f$, for the NPZ
dynamics without flow. As indicated in the legend, squares are for
$P$ and circles for $PP$.
\label{fig:tiempos}}
\end{figure}

To better understand this situation we have performed numerical
analysis of the NPZ dynamics without hydrodynamic terms. Figure
\ref{fig:tiempos} shows the time needed for $P$ and $PP$ to reach
their maximum values as a function of the initial conditions. We
define $f$ as the fraction of the ambient concentrations that are
present in the inflow: $(N,P,Z)=f\times(N_{amb},P_{amb},Z_{amb})$.
 With a mean flow of speed
$u_0=0.18 m/s$, fluid elements spend only $16$ days ($0.53$ in
units of $T_c$) inside the domain of horizontal extension $10r=250
km$ in figure \ref{fig:area}.  We see in Fig. \ref{fig:tiempos}
that for $f < 0.4$ the maximum in $P$ occurs later, so that we
cannot expect considerable growth in $A_s$ outside the vortices
for the value of $f=0.01$ used in  this work. But this observation
is also puzzling, since the primary production reported in figure
\ref{fig:PP-over-t} for the low inflow case is not as small as the
above argument would indicate: It is reduced just between $40$\%
and $85$\% with respect to  the ambient values.

Figure \ref{fig:patterns} (right column) clarifies the mechanisms
involved. The spatial plankton distribution is rather different
from the ambient inflow case. It is clearly related to the
vortices and associated structures. The plankton concentration is
very low outside these objects. As in the case of ambient inflow,
a plume with high nutrient concentration is present in the system
due to upwelling and has a shape similar to the one in the ambient
inflow case, which resembles the phytoplankton distribution of the
central column of fig.~\ref{fig:patterns},
 but here it seems to have no effect in inducing
phytoplankton growth. The time scale for plankton growth starting
from small values is much larger than the travel time through the
computational area, thus these effects are observable only further
downstream. Therefore, in the study area displayed in figures
\ref{fig:area} and \ref{fig:patterns} the influence of the
upwelling nutrient filament is masked by a more prominent
mechanism, described below. In fact, in this low-inflow case, we
note that the phenomenology observed in the area remains
qualitatively unchanged if the upwelling is removed, although
quantitative changes occur. This indicates that the dominant
mechanism in the low inflow situation is not the mixing of
upwelling and ambient waters, as in the ambient inflow situation,
but the interaction of the inflow with the wake.

\begin{figure}
\centering{ \mbox{
\includegraphics
[angle=0,width=\textwidth]{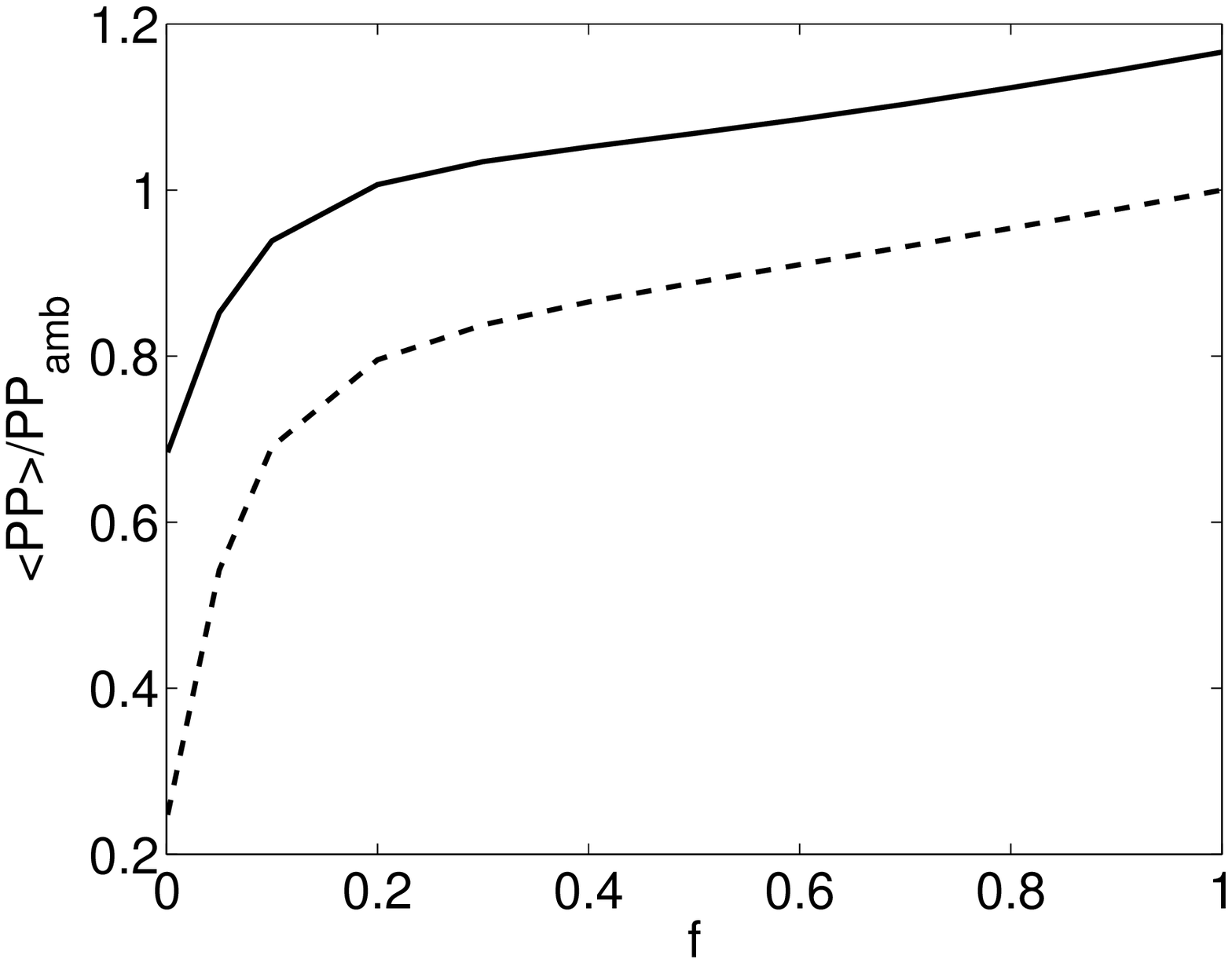} } }
\caption{The spatiotemporal average of the primary production in
$A_s$ in terms of its ambient value $<PP>/PP_{amb}$ as a function
of the inflow concentrations at vortex strength $w=w_H$ (solid
line) and $w=w_L$ (dashed line). $f$ is the fraction of the
ambient concentrations in the fluid entering the system.}
\label{fig:PP-over-inflow}
\end{figure}

Figure \ref{fig:patterns} shows that phytoplankton growth in the
vortices occurs after phytoplankton is transported into their
interior by filaments emerging from the boundary of the circular
obstacle. This complex structure -- boundary of the obstacle,
filaments emerging from it and rolling up around vortices -- is
well known from dynamical systems studies of this kind of flow
\citep{Jung1993,Ziemniak1994,Karolyi2000,Tel2005,Sandulescu2006},
and is related to the so-called {\sl unstable manifold of the
chaotic saddle}, the main dynamical structure in the wakes
occurring in time-dependent two-dimensional open flows. Loosely
speaking, it is the location of the fluid elements that take a
long time to leave the proximity of the island, because of the
complex recirculation emerging just behind the obstacle as well as
the reduced velocities occurring near its boundary. A detailed
analysis of these structures and their implications for the
residence times has been published in \cite{Sandulescu2007}.
Particle release calculations allow us to realize that, although
most of the incoming particles follow the mean flow and leave the
system in the $16$ days lapse estimated before, a fraction of them
are captured by the wake structures with residence times of about
$50$ to $20$ days for $w_H$ and  $w_L$, respectively. These long
residence times allow the plankton concentration to build up in
the filaments emerging from the obstacle, which gives rise to a
plankton bloom later downstream, when the filaments are stretched
and rolled up by the vortices.

Thus, recirculating structures in the island wake act as {\sl
incubators} that make fluid elements more productive before
releasing them into the main current. It turns out that the peak
values of the phytoplankton bloom in this low inflow case are
larger than the ones under ambient inflow. This somehow
paradoxical observation is explained by the fact that zooplankton
values are relatively high under equilibrium ambient conditions,
so that grazing control of the phytoplankton population is rather
effective. By contrast, in the low inflow case zooplankton and
thus grazing control is essentially absent. Zooplankton
concentration begins to build up only when phytoplankton
concentration has already reached larger values. Therefore it is
responsible for the end of the bloom further downstream, but high
phytoplankton values are attained before that.

Figure \ref{fig:PP-over-t} (two lower curves) shows the time
evolution of the primary production under the low inflow
conditions. Even for $w=w_L$, for which the wake acts as a barrier
blocking nutrient fertilization of $A_s$ from the upwelling,
primary production shows an oscillating behavior, reflecting the
oscillations of the wake structure which is the responsible for
the plankton growth. Figure \ref{fig:PP-over-w} shows the increase
in primary production by increasing the vortex strength $w$
starting at a value of $w=w_c \sim 0.1 w$. In the range of $w$
considered there is an increase in primary production by a factor
of about $2.17$, larger than the factor $1.2$ of increase attained
under ambient inflow.

So far we have described two very distinct inflow situations and
studied the impact of vortices by varying the vortex strength. We
now fix the vortex strength $w$ to the high, $w_H$, and low,
$w_L$, values,
 and describe the primary production
behavior for intermediate inflow cases in figure
\ref{fig:PP-over-inflow} (solid line is for $w_H$ and dashed line
for $w_L$). The inflow concentrations are now varied in terms of
$f$ which, as already mentioned, gives the fraction of the ambient
concentrations that are present in the inflow.
 For $f=0.01$ we have
the low inflow conditions considered before, and $f=1$ corresponds
to ambient input. We see, for both values of $w$, an  increase of
the average production in $A_s$ with the biological content of the
inflow, which is rather fast until $f\approx 0.2$.

\begin{figure}
\centering{ \mbox{
\includegraphics
[angle=0,width=\textwidth]{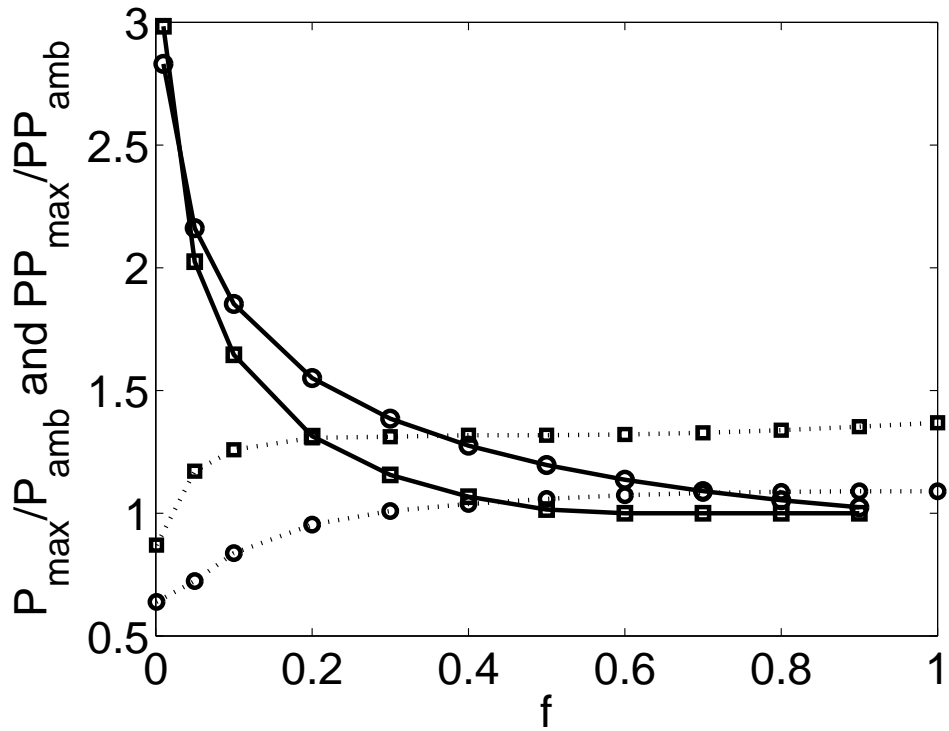} } }
\caption{Maxima of the average in $A_s$ of $P$ and $PP$ (normalized with the
corresponding ambient values) vs $f$. The circles are for the $P$
concentrations and the squares for $PP$. Data points joined with
solid line are computed without flow, and the ones joined by
dashed line are computed with flow. }
\label{fig:redistributing}
\end{figure}

Maximum values of the averages of $P$ and $PP$ in $A_s$ are shown
in Fig.\ref{fig:redistributing} for different initial conditions,
with and without flow (i.e. in the spatially homogeneous case). It
is seen a contrasting behavior remarking the fact that the flow is
not simply redistributing biological material, but that it
modifies also the spatial averages and then the total amounts of
substances, as well as the qualitative dependence on $f$. The
large overshoots of the homogeneous case for low $f$ only occur
inside vortices in the presence of flow, as will be discussed in
the next section, so that the spatially averaged values of $P$ and
$PP$ are lower (see also Fig. \ref{fig:PP-over-t}). When
approaching $f=1$ (the ambient inflow case), the relative increase
in the presence of flow arises by advection from the upwelling
plume.

\begin{figure}
\centering{ \mbox{
\includegraphics
[angle=0,width=\textwidth]{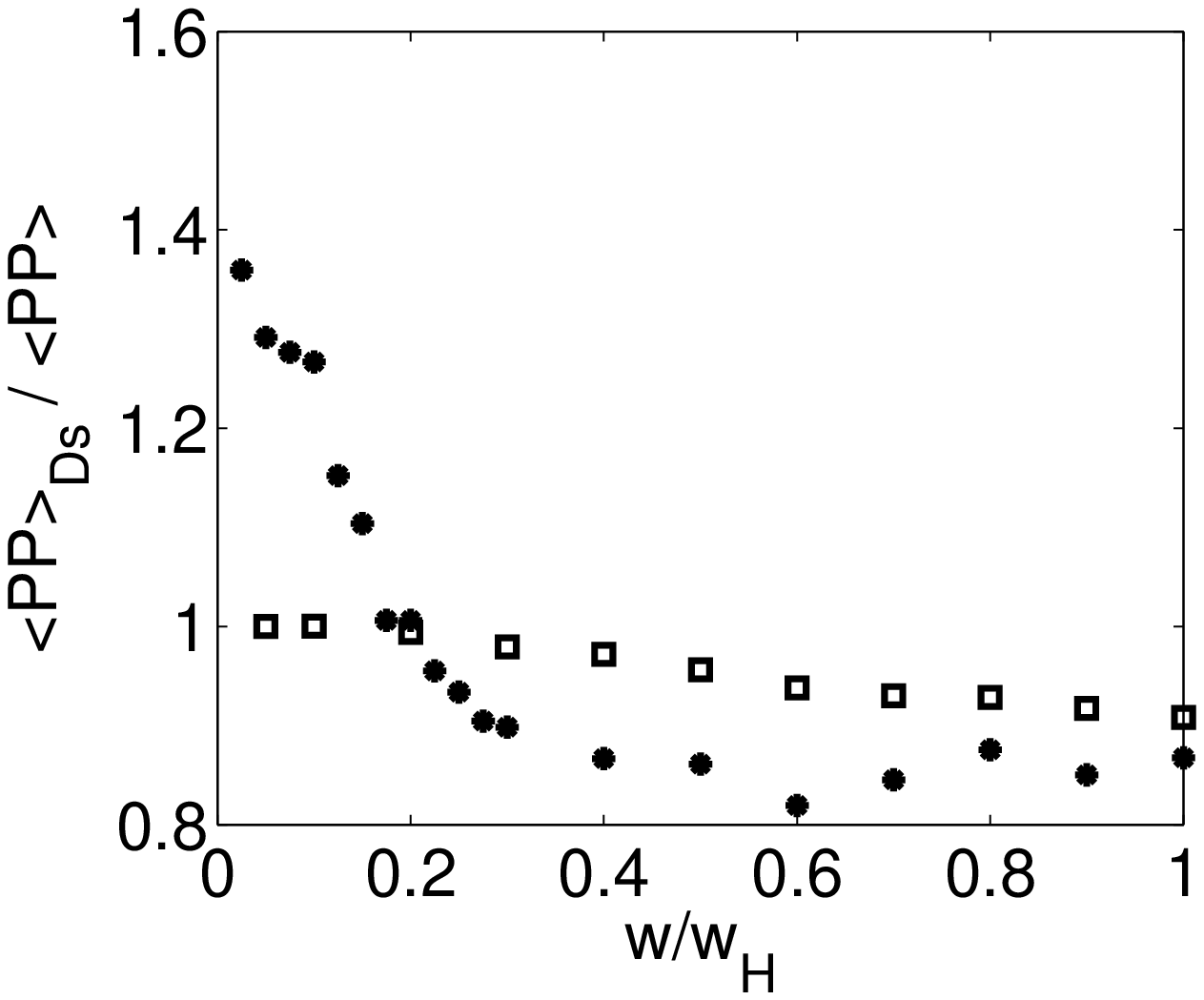} } }
\caption{
Average value of $PP$ in $A_S$ computed for a small value of the
diffusion coefficient, $D_s=0.1 D$, over the one obtained for $D$,
versus $w/w_H$. Filled circles correspond to the low inflow case,
while squares to the ambient inflow one. }
\label{fig:difusion}
\end{figure}

For completeness, we present in Fig.\ref{fig:difusion} the effect
of reducing the effective diffusion $D$. Physically this would
mean that we decrease the intensity of small scale turbulence. We
see that in general primary production is slightly reduced. The
increase which is observed at low values of $w$ in the low inflow
case is due to the lack of dilution of the filaments which emerge
from the cylinder boundary. For the ambient inflow situation a
small decrease of productivity for smaller $D$ is seen only at
large vorticity, because distributions in $A_s$ are rather
homogeneous otherwise.

\subsection{Vortices and plankton distribution}
\label{subs:Vortices}

It is well known that vortices are responsible for a large part of
the transport and mixing phenomena at mesoscale on the ocean
surface \citep{Barton1998,Pelegri2005,Martin2002}. They influence
biological dynamics, and most of the studies have focused on the
effect of the relatively large vertical motions induced by their
cores. Here we focus instead on horizontal processes. In this
section we consider the case $w=w_H$, where strong vortices are
present in the system, and characterize the plankton distributions
relative to vortex positions for the two different inflow
conditions, that highlight the two different primary production
enhancement mechanisms discussed above. Similar results are
expected for other values of $w$.

We make use of the Okubo-Weiss parameter
\citep{Okubo1970,Weiss1991} $W$ (a precise definition is included
in the Appendix~\ref{appendix:OW}) to identify in an objective way
the interior and the exterior part of vortices. Flow regions with
$W<0$ are vorticity dominated, and can be identified as the inner
part of vortices. Regions with $W>0$ are strain dominated and
outside vortices. The leftmost column of figure~\ref{fig:patterns}
displays the $W$ values, showing clearly the position of the
vortices. These positions can be correlated with the phytoplankton
distributions displayed in the other columns.

\begin{figure} 
\centering{ \mbox{
\includegraphics [angle=0, width=\textwidth]{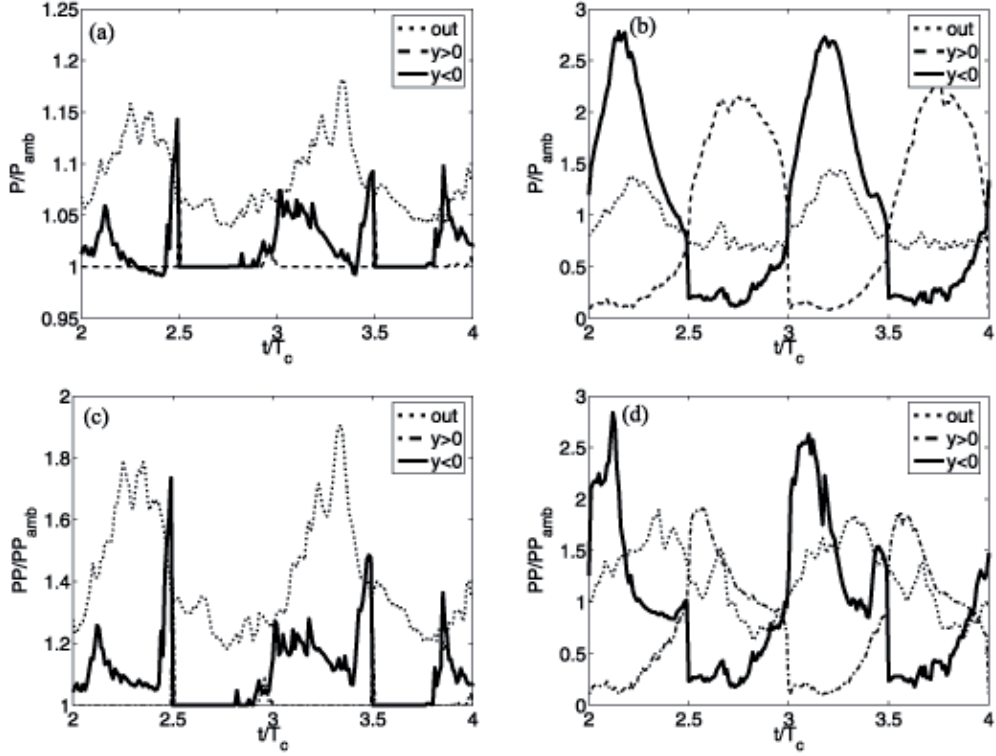}
} }\caption{Upper panels: time evolution of the spatial average of
the phytoplankton concentration $P$ (normalized with $P_{amb})$
inside the two vortices: $y<0$ (lower vortex, solid line) and
$y>0$ (upper vortex, dashed line), and out (outside, dot line).
Left is for ambient inflow concentrations, and right for low
inflow concentrations.  Vortex strength is always $w=w_H$. Lower
panels: primary production in the same locations and situations.
\label{fig:vortex-P-PP-over-t}}
\end{figure}

Figure \ref{fig:vortex-P-PP-over-t} shows the spatial average of
phytoplankton concentrations and primary production inside each of
the two vortices and outside them. These regions are identified
with the help of the Okubo-Weiss parameter $W$. The primary
production time series is qualitatively similar to the
phytoplankton one, although slightly shifted towards earlier
times. This is so because $PP$
 contains the influence of the nutrient dynamics, whose
temporal evolution anticipates the phytoplankton one. The
zooplankton time series (not shown) are also qualitatively similar
but shifted towards later times.

In the case of ambient inflow the interior of the vortices (dashed
and continuous lines) contains the same quantity of plankton as
the inflow, namely the ambient concentration. Only when additional
nutrients from the upwelling zone are entrained we observe bursts
localized in time. Thus most of the biological activity is in the
outside area (which includes the upwelling zone). This quantifies
what is seen in the middle column of figure \ref{fig:patterns}:
plankton appears mainly in filaments that wind around the vortex
periphery basically without entering them. The asymmetry observed
between the content of the two vortices arises from the fact that,
due to the different sense of rotation of the vortices more
nutrients are transported towards the vicinity of the lower vortex
than to the upper one (cf. figure \ref{fig:patterns}).

The situation is rather different in the low inflow case. The
range of the concentration oscillations is now larger, and the
content of the two vortices oscillates in antiphase. The largest
concentration values occur now inside vortex cores, leading to
peak bloom values larger than for the ambient inflow situation.
Minima are also smaller so that averages in regions such as $A_s$
give an overall smaller plankton content and primary production.
Filamental structures close to the boundary of the island
transport concentrations towards the vortices where the species
are trapped and transported downstream. During this motion their
concentrations are homogenized by small scale turbulence (modeled
by the diffusion term in Eq.~\ref{PDE}) and the classical dynamics
that the system of equations (\ref{BioDLG}) exhibits in a
homogeneous situation occurs: nutrient consumption by
phytoplankton induces a large phytoplankton bloom which is stopped
by the grazing by zooplankton, that also experiences growth, until
all three components approach the final  equilibrium value
$(N_{amb},P_{amb},Z_{amb})$. This steady state for the vortex
content occurs only further  downstream.

\section{Conclusions}
\label{conclusions}

We have presented numerical results on the biological dynamics in
the wake of an island close to a coastal upwelling area. Parameter
values were appropriate for the Canary Islands region but we
expect our results to be of greater generality.

Two different scenarios have been identified and discussed. In the
first one, occurring when the region outside the focus area has
properties similar to it, we have identified an enrichment
mechanism of one side of the wake by nutrients upwelled on the
other side. It occurs at sufficiently high vortex strength of the
vortices present in the wake. Vortices entrain water from one side
of the island in the form of filaments that are transported across
the wake. Filaments of this type are observed in satellite images
of the Canary area \citep{Barton1998,Pelegri2005}. When the vortex
strength is low, the wake acts rather as a barrier that blocks
transport from one side to the other. This scenario is a direct
translation of the behavior of passive tracers under similar flow
\citep{Sandulescu2006}. We have also observed that a large
decreasing of the eddy diffusion coefficient value has no relevant
role in the primary productivity, though in general this is
slightly reduced.

The second scenario becomes evident when the waters surrounding
the study area are biologically much poorer. Now fertilization by
the upwelling is not relevant, but we have identified a mechanism
for primary production increase in the wake: The large residence
times of some of the fluid particles in particular structures of
the island wake allow them to become enriched by the ambient
nutrient sources. Filaments from the wake structures feed this
enriched water into the vortices, and the nonequilibrium plankton
dynamics there leads to strong plankton blooms confined inside the
vortices. The biological significance of hydrodynamical structures
in the wake of obstacles has been recognized before
\citep{Karolyi2000,Scheuring2000,Tel2005}, but in these cases the
relevance was associated with their complex geometric structure
that allowed fine intertwining of filaments containing different
species or substances. The mechanism presented here seems to be
different and simply associated with large residence times in the
wake, leading to a kind of {\sl incubatory} effect. A carefull
study of the role of the different timescales present in the
system has been presented in \citet{Sandulescu2007}. We expect
this mechanism to be at work in many types of island wakes, even
if they are not associated with upwelling systems.

\section{Acknowledgements}

The authors thank T. T{\'e}l for many inspiring discussions. M.S.
and U.F. acknowledge financial support by the DFG grant FE
359/7-1. E.H-G. and C.L. acknowledge financial support from MEC
(Spain) and FEDER through projects CONOCE2 (FIS2004-00953) and
FISICOS (FIS2007-60327), and from the PIF project OCEANTECH from
CSIC. Both groups have benefited from a MEC-DAAD joint program.

\appendix
\section{Numerical Algorithm}
\label{appendix:algorithm}

The coupling of the biological and the hydrodynamic model yields
an advection-reaction-diffusion system. This system is numerically
solved by a semi-Lagrangian algorithm. The concentrations of $N$,
$P$ and $Z$ are represented on a grid of [500 x 300] points. The
three processes, advection, reaction and diffusion, are performed
sequentially as follows:

\begin{itemize}

\item Advection: Every point of the grid is integrated
backwards in time with the velocity field for a time step $dt$. In
this way we obtain the past position of the chosen point, which
typically is not on the grid.

\item Reaction: the concentration of $N$, $P$, or $Z$ at
the position obtained in the advection step is computed using a
bilinear interpolation of its corresponding nearest points on the
grid. Then these concentrations are the starting values for the
integration of the reaction dynamics forward-in-time for a time
step $dt$. In this way, we obtain the concentration of $N$, $P$
and $Z$ in the original grid point.

\item Diffusion: Once we have the values of the concentrations on the grid,
the diffusion step is performed following an Eulerian scheme.
However, the interpolation in the reaction part induces a
numerical diffusion of the order $ D_n\propto dx^2/dt$. It is
therefore desirable to ensure that the real diffusion, which is
given by the Okubo estimation \citep{Okubo1971}$D=10m^2/s$,~is
larger than this numerical diffusion. Moreover, the stability
condition for the Eulerian diffusion step requires that $D dt_d
/dx^2 < 1$, where $dt_d$ denotes the time step for the diffusion
part. These two inequalities are fullfilled if the diffusion time
step $dt_d$ is smaller than $dt$. According to this condition we
chose $dt_d=dt/10$ in our algorithm. This implies that the
algorithm makes ten steps of diffusion $dt_d$ after every step of
advection and reaction. Expressed in units of $T_c=30~days$ for
time and $r=25~km$ for space, the dimensionless numerical values
of the parameters are $dx=0.02$, $dt=0.01$, $dt_d=0.001$, and
$D=0.041472$.

\end{itemize}

\section{The Okubo-Weiss parameter}
\label{appendix:OW}

The Okubo-Weiss parameter $W$~\cite{Okubo1970,Weiss1991} is a
quantity used to distinguish areas in which the flow is dominated
by vorticity from those areas where the flow is strain dominated.
It is given by

\begin{equation}
\label{OW-def}
W=s^{2}_{n}+s^{2}_{s}-\omega^{2},
\end{equation}

where $s_{n}$, $s_{s}$  are the normal and the shear components of
strain, and $\omega$ is the relative vorticity of the flow defined
as:

\begin{equation}
\label{OW-comp-def}
s_{n}=\frac{\partial u}{\partial x}-\frac{\partial v}{\partial y},
~~~ s_{s}=\frac{\partial v}{\partial x}+\frac{\partial u}{\partial
y} ~~~, \omega=\frac{\partial v}{\partial x}-\frac{\partial
u}{\partial y}.
\end{equation}

Where $u$ and $v$ are the horizontal components of the velocity
field. We chose the critical threshold value  to be $W_{c}=0$. For
areas where $W$ is below $W_{c}$ the flow is vortex dominated,
otherwise we consider the flow to be strain dominated.

\end{document}